RESEARCH ARTICLE

# Efficient and accurate numerical-projection of electromagnetic multipoles for scattering objects

Wenfei Guo[1], Zizhe Cai[1], Zhongfei Xiong[1,*], Weijin Chen[2], Yuntian Chen[1,3,4,*]

[1]*School of Optical and Electronic Information, Huazhong University of Science and Technology, Wuhan 430074, China*

[2]*Department of Electrical and Computer Engineering, National University of Singapore, Singapore 117576, Singapore*

[3]*Wuhan National Laboratory of Optoelectronics, Huazhong University of Science and Technology, Wuhan 430074, China*

[4]*Optics Valley Laboratory, Wuhan 430074, China*

[*]*xiongzf94@outlook.com; yuntian@hust.edu.cn*

**Abstract**

In this paper, we develop an efficient and accurate procedure of electromagnetic multipole decomposition by using the Lebedev and Gaussian quadrature methods to perform the numerical integration. Firstly, we briefly review the principles of multipole decomposition, highlighting two numerical projection methods including surface and volume integration. Secondly, we discuss the Lebedev and Gaussian quadrature methods, provide a detailed recipe to select the quadrature points and the corresponding weighting factor, and illustrate the integration accuracy and numerical efficiency (that is, with very few sampling points) using a unit sphere surface and regular tetrahedron. In the demonstrations of an isotropic dielectric nanosphere, a symmetric scatterer, and an anisotropic nanosphere, we perform multipole decomposition and validate our numerical projection procedure. The obtained results from our procedure are all consistent with those from Mie theory, symmetry constraints, and finite element simulations.

## 1. Introduction

Electromagnetic multipoles play an indispensable role across different sub-branches of optics and photonics [1]. In classical electrodynamics, the field excited by localized spatial charge and current distribution can be decomposed into electric and magnetic multipole fields of all orders [2]. A visualization of field decomposition is presented in Fig. 1. The total scattered field can be decomposed into a series of multipole modes of all orders, including dipole, quadrupole, octupole, and high-order multipoles. In optics, the multipole decomposition method has been widely used in many nanophotonics scenarios for the analysis of light scattering of both a single scatterer and of periodic arrays of nanostructures [3-10]. The electric and magnetic multipoles excited in the scatterer contain valuable information about its optical response, including resonance, far-field patterns, and scattering cross-section [11-15]. Moreover, high-order multipoles such as electric and magnetic octupoles can be structurally engineered to tune optical resonances, absorption and scattering [16-18].

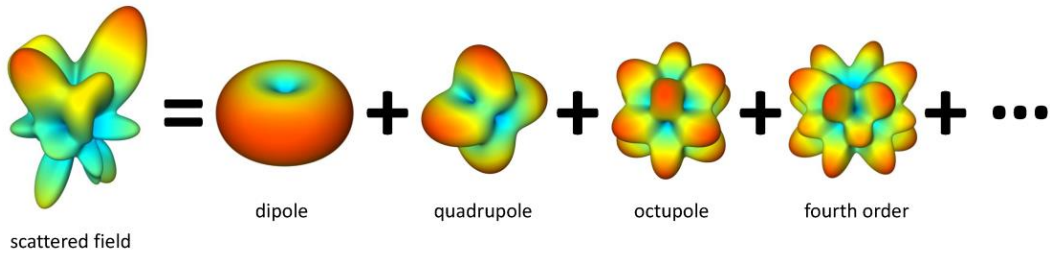

**Figure 1**. Schematic of multipole decomposition. The leftmost field pattern represents the total scattered field, which can be irregular and complex. The field patterns on the right show the multipole modes of order 1,2,3 and 4.

Apart from symmetric scatterers such as spheres or cylinders, normally there are no analytical solutions to the electromagnetic multipoles of irregular scatterers. Hence, numerical projection of electromagnetic multipoles is desirable for two purposes: (1) for understanding the underlying principles of many exotic phenomenon such as ultra-directional scattering, perfect reflection and transmission, anapole effects [19-24]; (2) for designing photonic devices such as metasurfaces and plasmonic arrays [25-31]. Several routes are feasible for realizing numerical projection of electromagnetic multipoles, as is well documented in recent works [32-34]. For instance, Alaee et al. proposed a multipole expansion method based on exact multipole moments beyond long-wavelength approximation [32]. Evlyukhin et al. studied discrete dipole approximation to calculate the multipole components of non-spherical scatterers [33]. Evlyukhin et al. also combined analytical and numerical methods to calculate the multipoles of scatterers with anisotropic optical properties [34]. The approach for multipole decomposition considered in this paper is based on multipole modes in spherical coordinates, rather than multipole moments in Cartesian coordinates [16,17,23,24,35]. Evidently, numerical integration is one of the key ingredients to perform multipole decomposition, and can be carried out using surface or volume integral techniques, provided that the scattered field is known from full-wave simulation. Despite the relevance of numerical integration in multipole decomposition, previous studies provide no systematic treatment of how the numerical integration is carried out, including the selection of sampling points, numerical accuracy and efficiency.

In this paper, we propose an efficient and accurate numerical projection procedure for multipole decomposition, which is complementary to the existing works in multipole projection. To the best of our knowledge, the Lebedev and Gaussian quadrature methods are introduced for the first time to process the numerical projection of multipoles, wherein the integration sampling points can be selected efficiently and accurately based on the solid framework given by the quadrature rule, i.e., in accordance with the interpolation functions used in full-wave simulations such as finite element method. Moreover, high-order electric and magnetic multipoles up to the 8th order are accurately coded in our program. The codes for MATLAB implementation of our new numerical projection procedure are open-source and available on GitHub [36].

The paper is organized as follows. In Section 2, we review the general approach to multipole decomposition for scattering problems. Then the Lebedev and Gaussian quadrature methods for surface and volume integration are introduced and tested in detail. In Section 3, we validate the numerical projection procedure using benchmarks of an isotropic dielectric nanosphere, a symmetric scatterer, and an anisotropic nanosphere. Finally, Section 4 summarizes the paper.

## 2. Theory

In this section, we briefly review the principles of the multipole decomposition method. Two types of numerical projection methods are highlighted, including surface integration based on the scattered field and volume integration based on the induced current density. To perform surface and volume integration in the numerical projection procedure, the Lebedev and Gaussian quadrature methods are adopted and discussed in detail.

The general working flow of the algorithm, following the common procedure of multipole decomposition based on numerical projection, is depicted in Fig. 2. Firstly, for a given structure, the scattered and total field in the whole space is computed using full-wave simulation such as finite element method (FEM) or finite-difference time-domain (FDTD) method. Based on the field data, the multipole coefficients of the scattered field are obtained by numerical projection. There are two approaches for numerical projection, each suitable for different situations. The first approach is surface integration based on the scattered field outside the scatterer, which is mainly used for the multipole analysis of a single structure. The second approach is volume integration based on the induced current density inside the scatterer, which is applicable to the building blocks in periodic structures such as metasurfaces and photonic crystals. Examples show that the surface integration method has faster convergence than the volume integration method with less computational resources in practice.

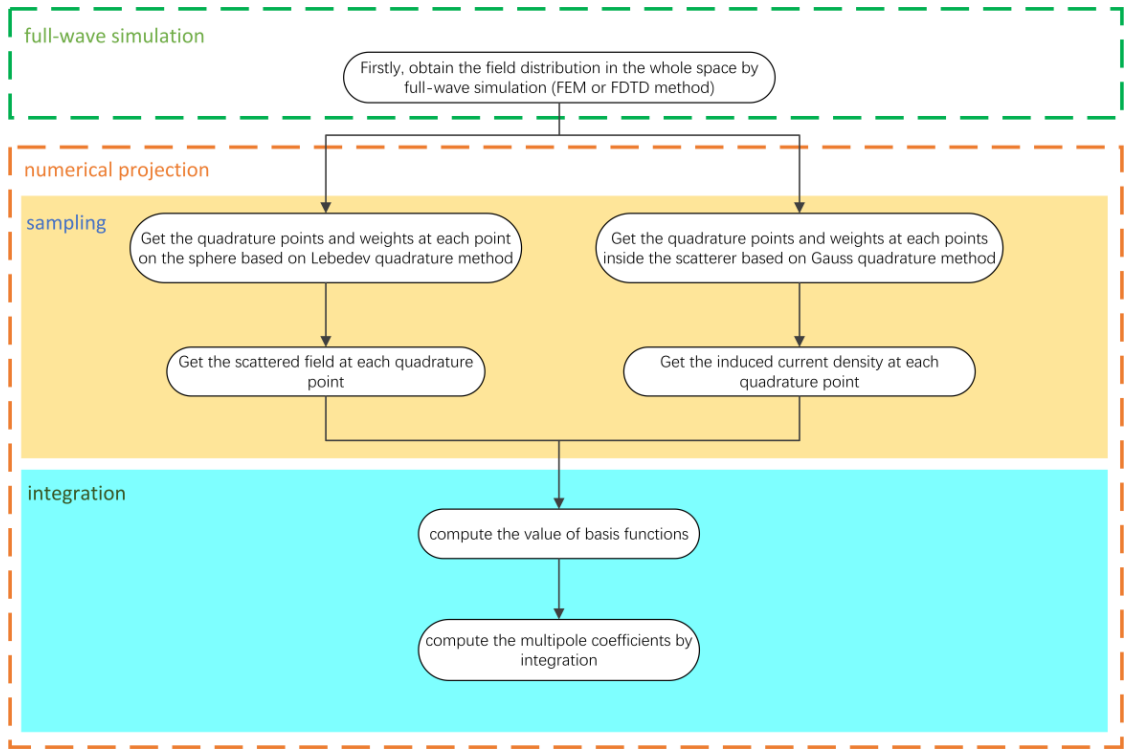

**Figure 2**. Working flow of the algorithm for multipole decomposition.

### 2.1. Surface integration based on the scattered field

The first approach to numerical projection of multipoles is based on the scattered field outside the scatterer. Consider a monochromatic incident plane wave $\boldsymbol{E}_i = \boldsymbol{E}_0 e^{i(\boldsymbol{k}\cdot\boldsymbol{x}-\omega t)}$, where $\omega$ is the angular frequency, $\boldsymbol{E}_0$ is the complex amplitude of electric field, and $\boldsymbol{k}$ is the wave vector in the background material. The total electric/magnetic field is denoted by $\boldsymbol{E}/\boldsymbol{H}$, which is the sum of the incident field $\boldsymbol{E}_i/\boldsymbol{H}_i$ and scattered field $\boldsymbol{E}_s/\boldsymbol{H}_s$, i.e., $\boldsymbol{E} = \boldsymbol{E}_i + \boldsymbol{E}_s$ and $\boldsymbol{H} = \boldsymbol{H}_i + \boldsymbol{H}_s$. In the source-free region, the wave equations for total electric and magnetic field are written as follows:

$$\nabla^2 \boldsymbol{E} + k^2 \boldsymbol{E} = 0,$$
$$\nabla^2 \boldsymbol{H} + k^2 \boldsymbol{H} = 0. \qquad (1)$$

The wave vector $k = \sqrt{\boldsymbol{k} \cdot \boldsymbol{k}} = \omega\sqrt{\varepsilon\mu}$, where $\varepsilon$ and $\mu$ are the permittivity and permeability of isotropic background. Adapting the spherical coordinate system, the eigen solutions of wave equations Eq. (1) are vector spherical harmonics. The concrete forms of vector spherical harmonics $\boldsymbol{M}$ and $\boldsymbol{N}$ of order $(l, m)$ are expressed as follows [37]:

$$\boldsymbol{M}_{lm} = [i\pi_{lm}(\theta)\hat{\boldsymbol{e}}_\theta - \tau_{lm}(\theta)\hat{\boldsymbol{e}}_\phi] z_l(kr) e^{im\phi},$$
$$\boldsymbol{N}_{lm} = l(l+1) P_l^m(\cos\theta) \frac{z_l(kr)}{kr} e^{im\phi} \hat{\boldsymbol{e}}_r + [\tau_{lm}(\theta)\hat{\boldsymbol{e}}_\theta + i\pi_{lm}(\theta)\hat{\boldsymbol{e}}_\phi] \frac{[krz_l(kr)]'}{kr} e^{im\phi}, \qquad (2)$$

where $P_l^m(\cos\theta)$ is the associated Legendre polynomial, $z_l(kr)$ is the spherical Bessel function, and $[krz_l(kr)]'$ denotes the differentiation with respect to the argument $kr$. The auxiliary functions $\tau_{lm}(\theta)$ and $\pi_{lm}(\theta)$ are defined as:

$$\tau_{lm}(\theta) = \frac{d}{d\theta} P_l^m(\cos\theta),$$
$$\pi_{lm}(\theta) = \frac{m}{\sin\theta} P_l^m(\cos\theta). \qquad (3)$$

In physics, $\boldsymbol{N}_{lm}$ and $\boldsymbol{M}_{lm}$ represent the field excited by electric and magnetic multipoles, respectively, located at the origin in the spherical coordinate system. Because of the completeness of vector spherical harmonics, the scattered field $\boldsymbol{E}_s$ can be expanded in terms of the electric and magnetic multipole modes,

$$\boldsymbol{E}_s = \sum_{l=1}^{\infty} \sum_{m=-l}^{l} E_{lm} (a_{lm} \boldsymbol{N}_{lm} + b_{lm} \boldsymbol{M}_{lm}). \qquad (4)$$

where $a_{lm}$ and $b_{lm}$ are electric/magnetic multipole coefficients, which represent the contributions of electric/magnetic multipoles to the scattered field. The constant coefficient $E_{lm}$ is given by: $E_{lm} = \frac{i^{l+1}(2l+1)}{2} \sqrt{\frac{(l-m)!}{l(l+1)(l+m)!}}$. Considering the orthogonality of vector spherical harmonics, the multipole coefficients $a_{lm}$ and $b_{lm}$ are computed as follows [37]:

$$a_{lm} = \frac{\iint \boldsymbol{N}_{lm}^* \cdot \boldsymbol{E}_s \mathrm{d}s}{E_{lm} \iint \boldsymbol{N}_{lm}^* \cdot \boldsymbol{N}_{lm} \mathrm{d}s},$$
$$b_{lm} = \frac{\iint \boldsymbol{M}_{lm}^* \cdot \boldsymbol{E}_s \mathrm{d}s}{E_{lm} \iint \boldsymbol{M}_{lm}^* \cdot \boldsymbol{M}_{lm} \mathrm{d}s}, \quad (5)$$

where the asterisk $*$ denotes complex conjugation. The surface integration in Eq. (5) is taken over a spherical surface which encloses the scattering object. In practice, the scattered field $\boldsymbol{E}_s$ is computed by full-wave simulation, and the integration surface should not go beyond the simulation region.

**2.2. Volume integration based on the induced current density**

The second approach to numerical projection of multipoles is based on the induced current density inside the scatterer, which is defined as follows:

$$\boldsymbol{J}_s(\boldsymbol{r}) = -\mathrm{i}\omega\varepsilon_0(\varepsilon_r - \varepsilon_b)\boldsymbol{E}(\boldsymbol{r}), \quad (6)$$

where $\varepsilon_r$ is the permittivity distribution in the whole space with presence of the scatterer, $\varepsilon_b$ is the permittivity of background material, and $\boldsymbol{E}(\boldsymbol{r})$ is the total electric field. Based on Maxwell equations, the relations between the scattered field $\boldsymbol{E}_s/\boldsymbol{H}_s$ and the induced current density $\boldsymbol{J}_s$ are expressed as follows:

$$\begin{aligned}
\nabla \cdot \boldsymbol{E}_s &= -\frac{\mathrm{i}\eta}{k}\nabla \cdot \boldsymbol{J}_s, \\
\nabla \cdot \boldsymbol{H}_s &= 0, \\
\nabla \times \boldsymbol{E}_s &= \mathrm{i}k\eta \boldsymbol{H}_s, \\
\nabla \times \boldsymbol{H}_s &= -\frac{\mathrm{i}k}{\eta}\boldsymbol{E}_s + \boldsymbol{J}_s.
\end{aligned} \quad (7)$$

The wave equations for the scattered field $\boldsymbol{E}_s$ can be derived based on Eq. (7). Then one can further determine the exact forms of the induced current density of electric and magnetic multipoles of order $(l, m)$ [38],

$$\begin{aligned}
\boldsymbol{S}_{lm} &= [\Pi_l(kr) + \Pi_l''(kr)]P_l^m(\cos\theta)\mathrm{e}^{-\mathrm{i}m\phi}\hat{\boldsymbol{e}}_r + [\tau_{lm}(\theta)\hat{\boldsymbol{e}}_\theta - \mathrm{i}\pi_{lm}(\theta)\hat{\boldsymbol{e}}_\phi]\frac{\Pi_l'(kr)}{kr}\mathrm{e}^{-\mathrm{i}m\phi}, \\
\boldsymbol{T}_{lm} &= [\mathrm{i}\pi_{lm}(\theta)\hat{\boldsymbol{e}}_\theta + \tau_{lm}(\theta)\hat{\boldsymbol{e}}_\phi]\mathrm{e}^{-\mathrm{i}m\phi}J_l(kr),
\end{aligned} \quad (8)$$

where $\Pi_l(kr) = krJ_l(kr)$ is the Riccati-Bessel function. As in the case of the multipole decomposition based on the scattered field, one can also calculate the multipole coefficients based on the induced current density in the following form,

$$\begin{aligned} a_{lm} &= \frac{(-\mathrm{i})^{l-1}k^2\eta Q_{lm}}{\sqrt{\pi(2l+1)}}\iiint_V \boldsymbol{J}_s(\boldsymbol{r})\cdot\boldsymbol{S}_{lm}\mathrm{d}\Omega, \\ b_{lm} &= \frac{(-\mathrm{i})^{l+1}k^2\eta Q_{lm}}{\sqrt{\pi(2l+1)}}\iiint_V \boldsymbol{J}_s(\boldsymbol{r})\cdot\boldsymbol{T}_{lm}\mathrm{d}\Omega, \end{aligned} \quad (9)$$

where $Q_{lm} = \frac{1}{\sqrt{l(l+1)}}\sqrt{\frac{2l+1}{4\pi}\frac{(l-m)!}{(l+m)!}}$ is a constant coefficient and $\eta = \sqrt{\frac{\mu_0}{\varepsilon_0}}$ is the vacuum impedance. The two integrals in Eq. (9) are evaluated over the volume $V$ of the scattering object. The volume integration method applies to either a single scatterer or the building blocks in a periodic array. Since the unit building block expands in the entire full-wave simulation region in the direction with periodic boundary conditions, only the volume integration method, which extracts the induced current density inside the unit building block, can be used to calculate the multipole components.

It is worth mentioning that the multipole coefficients obtained using either the scattered field-based or induced current density-based numerical projections should be identical for a given scatterer. The expressions for the two methods can be transformed into each other using the relationship between the source current and far-field [38].

As a side remark, if the center of mass of the scatterer is not the same as the coordinate origin, then the multipole decomposition is complicated. Notably, the numerical projection procedures, i.e., surface or volume integration-based numerical projection as used in this paper, still apply and the resulting multipole components are not the same. However, the observable physical quantity obtained by the combination of different kinds of multipoles should be the same. If multipole decomposition is performed with respect to the symmetric center of the scatterer, there are fewer multipole components. For example, for a spherical scatterer, if the sphere center and the coordinate origin are identical, the projected multipole components have the fewest terms. In contrast, if the sphere center has an off-set from the coordinate origin, one can still do multipole decomposition. In such a case, there are more multipole components than is the case in the overlapped scenario, but the observable physical quantities are the same. For calculation of multipoles of asymmetrical or compound scatterers, it is beneficial to choose the center of mass as the reference point to get minimum number of multipole components.

### 2.3. Numerical implementations of integration

To compute the multipole coefficients accurately, certain numerical quadrature rules are needed to implement the surface or volume integration described in the above sections. In order to achieve higher computational accuracy with fewer function evaluations, we first adopt the Gaussian and Lebedev quadrature methods to compute the volume and surface integration in the numerical projection, respectively.

The Gaussian quadrature method is widely used in the numerical integration in the practical implementation of electromagnetic finite element computation. As such, the integration sampling points as well as the interpolation functions used in finite element full-wave simulation are also carefully selected according to the quadrature rule and can be directly used in our proposed method for multipole projection, wherein the integration sampling points can be cut down to the minimum, as constrained by finite element full-wave simulation.

### 2.3.1. Gaussian quadrature method for volume integration

Since the boundary of the integration region can be irregular, finite elements are often used to mesh the integration region. The most commonly used shape for meshing is the tetrahedron in three-dimensional space and the triangle for two-dimensional surfaces. To compute the integral over a tetrahedron, the Gaussian quadrature method is adopted due to its high accuracy. For a single tetrahedron $V$ with a unit volume, the numerical quadrature of a function $f$ over the tetrahedron $V$ is in the following form [39],

$$\iiint_V f(\alpha_1, \alpha_2, \alpha_3, \alpha_4) d\Omega \approx \sum_i \omega_i f(\alpha_1^i, \alpha_2^i, \alpha_3^i, \alpha_4^i), \tag{10}$$

where $\alpha_j^i (j = 1,2,3,4)$ are the natural coordinates of the $i$th quadrature point in the tetrahedron, and $\omega_i$ is the corresponding weight of the $i$th quadrature point. According to the definition of natural coordinates [40], the constraint for the quadruplet $(\alpha_1^i, \alpha_2^i, \alpha_3^i, \alpha_4^i)$ is $\sum_{j=1}^{4} \alpha_j^i = 1$. Because of the symmetry of tetrahedron, all possible permutations of the quadruplet $(\alpha_1^i, \alpha_2^i, \alpha_3^i, \alpha_4^i)$ are equivalent quadrature points sharing the same weight. The coordinates of the quadrature points and weights for the quadrature expression Eq. (10) are determined by integrating polynomials of particular forms exactly, leading to a set of nonlinear equations which are relatively difficult to solve [39]. The choice of polynomial order and form for exact integration depends on the desired degree $N$ of the quadrature method. For the integration over a tetrahedron, Gaussian quadrature formulas of degree $N = 4$ to 8 are given in Ref. [41].

To illustrate the Gaussian quadrature intuitively, we present an example of computing the integration of the function $f = x^2 + y^2 + z^2$ over a regular tetrahedron with a unit volume with only 11 sampling points. The geometry of the tetrahedron and selected quadrature points are shown in Fig. 3. The exact coordinates of these quadrature points and the corresponding weights are listed in Table 1. We utilize the Gaussian quadrature method with degree $N = 4$ for the computation, resulting in an integral value of 0.485352892045444 up to 16 digits of accuracy, which precisely matches the analytical solution of $\frac{7}{10 \times \sqrt[3]{3}} = 0.485352892045444$.

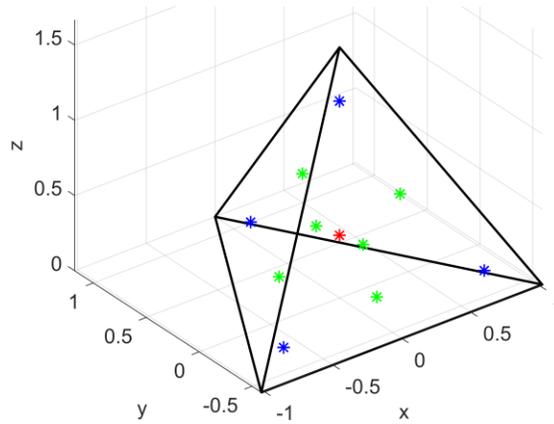

**Figure 3**. Geometry of the integration region and selected Gaussian quadrature points.

**Table 1**. The natural/Cartesian coordinates and weights of Gaussian quadrature points of degree $N = 4$ for the computation of integrals over a regular tetrahedron.

| Natural coordinates | Permutations | Weights | Cartesian coordinates |
|---|---|---|---|
| (0.25, 0.25, 0.25, 0.25) | 1 | -0.0789 | (0, 0, 0.4163) |
| (0.7857, 0.0714, 0.0714, 0.0714) | 4 | 0.0457 | (0.7284, -0.4206, 0.1190) |
| | | | (-0.7284, -0.4206, 0.1190) |
| | | | (0, 0, 1.3085) |
| | | | (0, 0.8411, 0.1190) |
| (0.3994, 0.3994, 0.1006, 0.1006) | 6 | 0.1493 | (0, 0.3519, 0.6652) |
| | | | (0.3047, -0.1759, 0.6652) |
| | | | (0.3047, 0.1759, 0.1675) |
| | | | (0, -0.3519, 0.1675) |
| | | | (-0.3047, 0.1759, 0.1675) |
| | | | (-0.3047, -0.1759, 0.6652) |

### 2.3.2. Lebedev quadrature method for surface integration

As mentioned above, Gaussian quadrature is compatible with finite elements such as tetrahedrons or triangles. It is straightforward to adapt Gaussian quadrature to compute the surface integration over a sphere, as long as the quadrature points and weights are computed based on the symmetry of triangles. However, note that a sphere has a higher degree of symmetry than a triangle, resulting in more quadrature points that are equivalent to each other. For instance, quadrature points which are equivalent under the octahedron rotation and inversion on the sphere share the same weight [42]. Therefore, a smaller number of independent quadrature points can achieve the same accuracy as Gaussian quadrature. This efficient quadrature method for a sphere is called Lebedev quadrature, which is adopted in this paper.

Given a unit sphere $S^2$ which is described by $x^2 + y^2 + z^2 = 1$, the integration of a function $f$ on $S^2$ is approximated using Lebedev quadrature in the following form [43],

$$I = \iint_{S^2} f(\mathbf{r}) d\sigma \approx A_1 \sum_{i=1}^{6} f(a_i^1) + A_2 \sum_{i=1}^{12} f(a_i^2) + A_3 \sum_{i=1}^{8} f(a_i^3) \\ + \sum_{k=1}^{N_1} B_k \sum_{i=1}^{24} f(b_i^k) + \sum_{k=1}^{N_2} C_k \sum_{i=1}^{24} f(c_i^k) + \sum_{k=1}^{N_3} D_k \sum_{i=1}^{48} f(d_i^k), \quad (11)$$

where the quadrature points $a_i^k$, $b_i^k$, $c_i^k$ and $d_i^k$ are determined according to the octahedronal symmetry, and $A_k$, $B_k$, $C_k$ and $D_k$ are the corresponding weights. To be more specific, for each group of quadrature points with the same $k$, they are invariant under the octahedral rotation group with inversion $G_8^*$ and share the same weight [43]. To determine the corresponding weights of quadrature points in Lebedev quadrature of order $n$, the quadrature method must exactly integrate over all spherical harmonics $Y_{lm}$ of order $l \leq n$. In addition, it is sufficient to determine the quadrature weights by requiring the exact integration of the polynomials that are invariant under the octahedral rotation with inversion. As in the case of the determination of Gaussian quadrature, the exact integration of polynomials or spherical harmonics also leads to a set of nonlinear equations to be solved. The computed value of weights $A_k,...,D_k$ and quadrature points for order $9 \leq n \leq 17$ can be found in Ref. [42].

Without loss of generality, we demonstrate an example of accurate integration for a function $f = x^2 + y^3 - z^4$, which is randomly chosen here, over a unit sphere using the Lebedev quadrature method with only 26 sampling points. The sampling points are shown in Fig. 4, and the coordinates and the corresponding weights are also listed in Table 2. Remarkably, the computed integral yields a value of 1.675516081914556, which shows agreement with the analytical solution of $\frac{8\pi}{15} = 1.675516081914556$, up to 16 digits of accuracy.

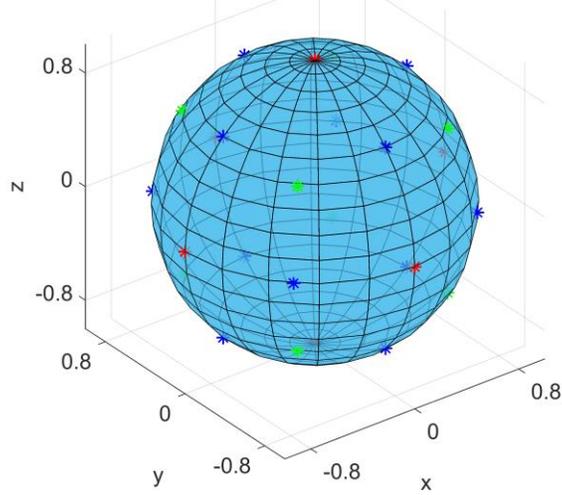

**Figure 4**. Geometry of the integration region and selected Lebedev quadrature points.

**Table 2**. The Cartesian coordinates and weights of Lebedev quadrature points of order $n = 7$ on a unit sphere.

| Cartesian coordinates | Permutations | Weights |
| --- | --- | --- |
| $(\pm 1, 0, 0)$ | 6 | 0.5984 |
| $(\pm 0.7071, 0, \pm 0.7071)$ | 12 | 0.4787 |
| $(\pm 0.5774, \pm 0.5774, \pm 0.5774)$ | 8 | 0.4039 |

## 3. Results and discussion

In this section, we study three scattering objects, including an isotropic dielectric nanosphere, a symmetric scatterer with $D_{3h}$ symmetry, and an anisotropic nanosphere, to validate our numerical projection procedure for multipole decomposition. The scattered field and induced current density are computed by full-wave simulation using COMSOL Multiphysics [44].

### 3.1. Scattering by an isotropic dielectric nanosphere

We compute the multipole scattering cross-section of an isotropic dielectric nanosphere as discussed in Ref. [45]. The nanosphere and the incident condition are illustrated in Fig. 5(a). The center of the nanosphere is located at the origin of the coordinate system. The nanosphere has a refraction index of $n = 3.5$ and a radius of $a = 210$ nm. The incident light is a plane wave propagating along the $z$ axis with linear polarization along the $x$ axis. The wavelength of incident light ranges from 800 to 1900 nm. For an isotropic spherical scatterer, Mie theory provides analytical expressions for the calculation of multipole coefficients [1]. Furthermore, we utilize both the surface and volume integration methods of numerical projection to compute the multipole coefficients. Based on the multipole coefficients $a_{lm}$ and $b_{lm}$, the total scattering cross-section $\sigma_{sca}$ is determined by the following expression,

$$\sigma_{sca} = \frac{2}{k^2 a^2} \sum_{l=1}^{\infty} \sum_{m=-l}^{l} (2l+1)\,(|a_{lm}|^2 + |b_{lm}|^2), \tag{12}$$

where $\sigma_{sca}$ is normalized based on the cross-sectional area $\pi a^2$ of the sphere. As shown in Fig. 5(a), it is evident that the numerical results from the surface and volume integration methods of our numerical projection procedure both have excellent agreement with the analytical results from Mie theory. The two resonance peaks are characterized accurately in the wavelength range from 900 to 1900 nm. Furthermore, it can be found that within this wavelength range, only the electric/magnetic dipole and quadrupole make significant contributions to the total scattering cross-section. In contrast, as shown in Fig. 5(b), the magnetic octupole has a notable impact in the wavelength range from 800 to 900 nm. In particular, the resonance peak at 818 nm is exclusively due to the magnetic octupole.

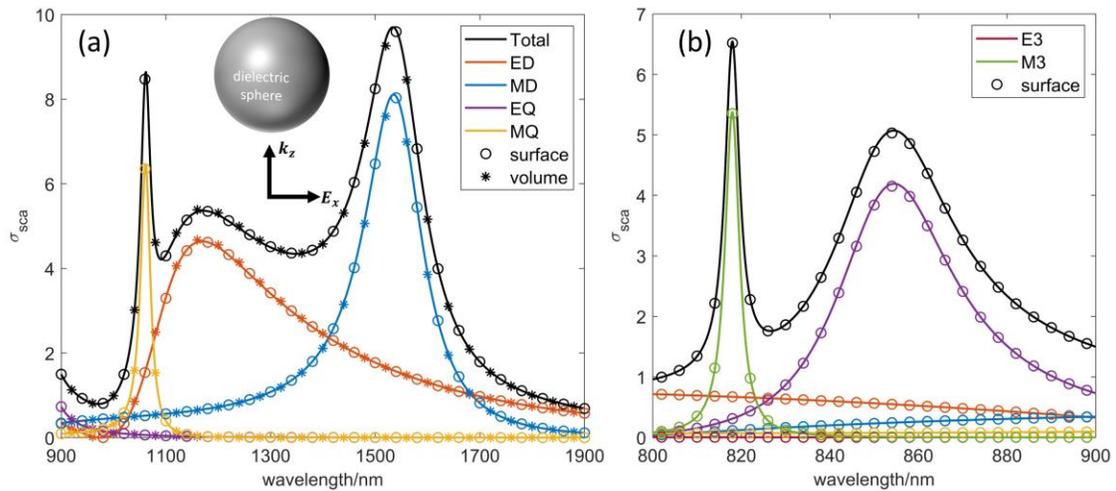

**Figure 5.** The multipole scattering cross-section of the dielectric nanosphere as a function of the incident wavelength. The analytical curves are derived from Mie theory, while the discrete symbols represent the numerical results from our procedure. Both surface and volume integration methods are utilized in subplot (a), while subplot (b) only employs surface integration.

The convergence analysis of the numerical projection procedure based on both surface and volume integration is shown in Fig. 6. The convergence of magnetic octupole (MO) scattering cross-section at incident wavelength of 818 nm is calculated using surface integration, and the convergence of magnetic quadrupole (MQ) scattering cross-section at 1000 nm is calculated using volume integration. The green region in Fig. 6 shows the convergence range for each case. The convergence thresholds are set to 1% and 0.2% for subplots (a) and (b), respectively. Evidently, Lebedev quadrature has a faster convergence than the trapezoidal rule of surface integration when the number of quadrature points is higher. Similarly, Gaussian quadrature also shows a faster and more stable convergence, while the trapezoidal rule of volume integration shows oscillation near the analytical value from Mie theory. It is also evident that the surface integration method requires fewer quadrature points than are required by the volume integration method. This is because meshing the volume of scatterer needs more finite elements than meshing the scatterer's surface.

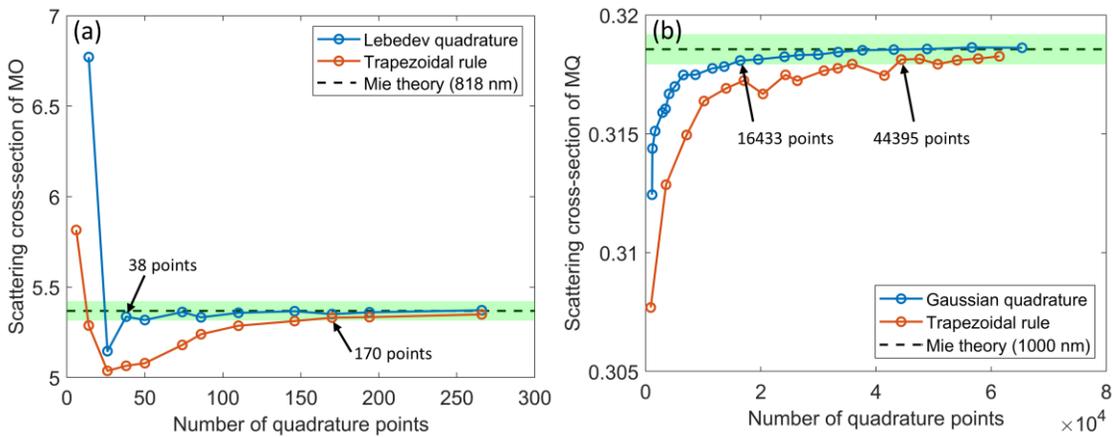

**Figure 6.** Convergence versus the number of quadrature points. Subplot (a) shows the convergence of magnetic octupole at 818 nm using surface integration. Subplot (b) shows the convergence of magnetic quadrupole at 1000 nm using volume integration. Black dashed line shows the analytical value from Mie theory.

As a side remark, the distribution of Lebedev quadrature points in this demonstration is the same as in the benchmark case (a unit sphere) in Sec. 2.3.2, up to a scaling factor. This is because Lebedev quadrature points are only determined by the Lebedev quadrature rule, regardless of the geometrical shape of the scatterer inside the integration sphere. However, to calculate volume integrals, the 3D scatterer needs to be meshed using tetrahedrons, so that the quadrature points are chosen at each tetrahedron. The Gaussian quadrature rule provides the quadrature points of different order in a reference tetrahedron. To apply Gaussian quadrature to real physical tetrahedron meshes, whose geometrical shape is generally irregular, there exists a mapping from the reference tetrahedron to physical tetrahedron. Therefore, the distribution of Gaussian quadrature points in each tetrahedron mesh for scattering particle cases can be different from the benchmark case (a regular tetrahedron) in Sec. 2.3.1, and the quadrature points can be transformed to each other by a mapping.

### 3.2. Scattering by a symmetric scatterer with $D_{3h}$ symmetry

The second benchmark evaluates the multipole coefficients of eigenmodes of a symmetric scatterer with $D_{3h}$ symmetry. In optics, symmetry can be used to classify eigenmodes in structures such as waveguides or photonic crystals [46]. Furthermore, optical symmetries can introduce certain constraints on the optical properties of the structure. For symmetric scatterers, a group theory approach can be applied to analyze the constraints of multipole coefficients imposed by the symmetry of the scatterer [47,48]. There are six irreducible representations of $D_{3h}$ group, and each eigenmode belongs to one of these representations. Specifically, we focus on two eigenmodes belonging to different representations $A_1''$ and $A_2''$. It is demonstrated that there are exact relationships between multipole coefficients of the eigenmodes. The constraints on the multipole coefficients can vary for eigenmodes belonging to different representations of the symmetry group. The character tables for the $A_1''$ and $A_2''$ repersentations, along with the corresponding constraints on the multipole coefficients of each representation are listed in Table 3. A detailed explanation of the symmetry operations in the $D_{3h}$ group can be found in Ref. [47].

**Table 3.** The character table of $D_{3h}$ group representation $A_1''$ and $A_2''$, and constraints on multipoles coefficients of the eigenmodes of two representations. $N$ is an integer.

| $D_{3h}$ | $E$ | $\sigma_h$ | $2C_3$ | $2s_3$ | $3C_2'$ | $3\sigma_v$ | Electric multipoles | Magnetic multipoles |
|---|---|---|---|---|---|---|---|---|
| $A_1''$ | 1 | -1 | 1 | -1 | 1 | -1 | $m = 3N$, $(l+m)$ is odd, $a_{l,-m} = (-1)^{m+1} a_{l,m}$ | $m = 3N$, $(l+m)$ is even, $b_{l,-m} = (-1)^m b_{l,m}$ |
| $A_2''$ | 1 | -1 | 1 | -1 | -1 | 1 | $m = 3N$, $(l+m)$ is odd, $a_{l,-m} = (-1)^m a_{l,m}$ | $m = 3N$, $(l+m)$ is even, $b_{l,-m} = (-1)^{m+1} b_{l,m}$ |

The multipole coefficients for the two eigenmodes in Fig. 7(a) and (c) are computed using the surface integration method of our numerical projection procedure. The results are shown in Fig. 7(b) and (d), respectively. The scatterer, represented by the solid black lines in Fig. 7(a) and (c), has a refractive index of $n = 8$ and is a three-dimensional structure, and only the $xy$ plane is depicted. It can be checked that both eigenmodes are invariant under the operation $E$ and $2C_3$, and are reversed under the operation $\sigma_h$ and $2s_3$. For the eigenmode shown in Fig. 7(a), the mode profile is reversed under the operation $3\sigma_v$, which represents reflection in the $xy$ plane. Therefore, this eigenmode satisfies the character of the $A_1''$ representation. The constraints listed in Table 3 for the $A_1''$ representation require the multipole coefficients to satisfy $b_{3,3} = -b_{3,-3}$ and $a_{4,3} = a_{4,-3}$, which perfectly match the computed multipole coefficients shown in Fig. 7 (b). Similarly, for the eigenmode in Fig. 7(c), the mode profile is unchanged under the operation $3\sigma_v$, indicating that this eigenmode satisfies the requirement of the $A_2''$ representation. The corresponding constraints for the $A_2''$ representation require $b_{3,3} = b_{3,-3}$ and $a_{4,3} = -a_{4,-3}$, which also agree with our numerical results as shown in Fig. 7(d).

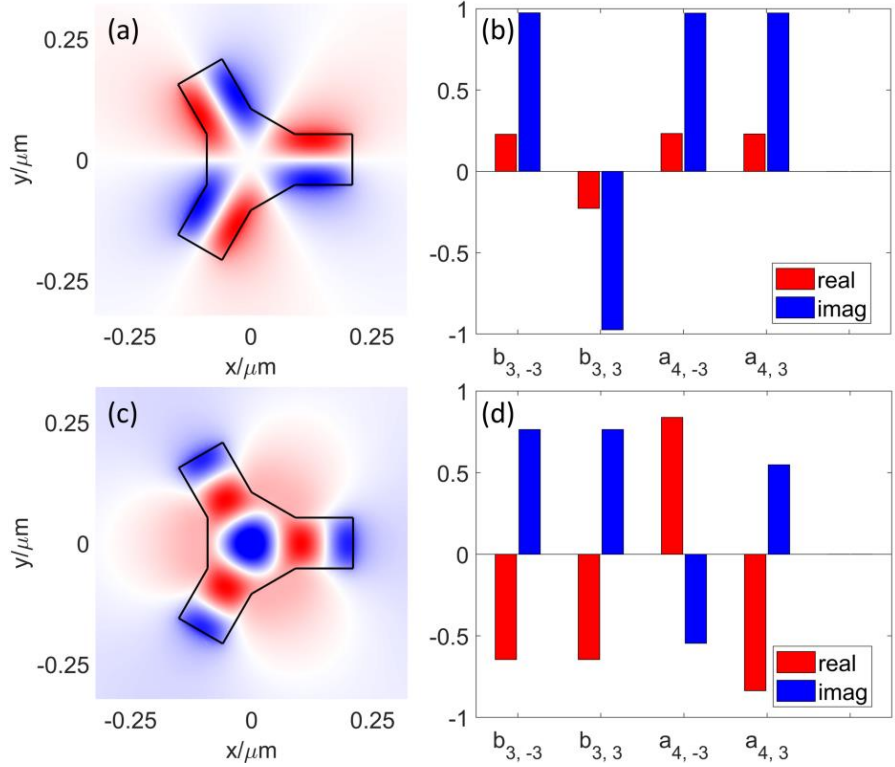

**Figure 7.** Two eigenmodes of the $D_{3h}$ scatterer and the normalized (with respect to itself) multipole coefficients of each eigenmode. The eigenmode shown in subplot (a)/(c) belongs to $A_1''/A_2''$ representation, whose multipole coefficients are presented in subplot (b)/(d).

As shown in Fig. 8, we evaluate the expression $\frac{b_{3,-3}+b_{3,3}}{b_{3,-3}}$ of the $A_1''$ eigenmode to analyze the convergence of surface integration-based numerical projection based on this benchmark. According to the symmetry constraints for $A_1''$ eigenmodes, the expression $\frac{b_{3,-3}+b_{3,3}}{b_{3,-3}}$ should be exactly zero. Evidently, Lebedev quadrature also has a faster convergence than the trapezoidal rule of surface integration in this benchmark.

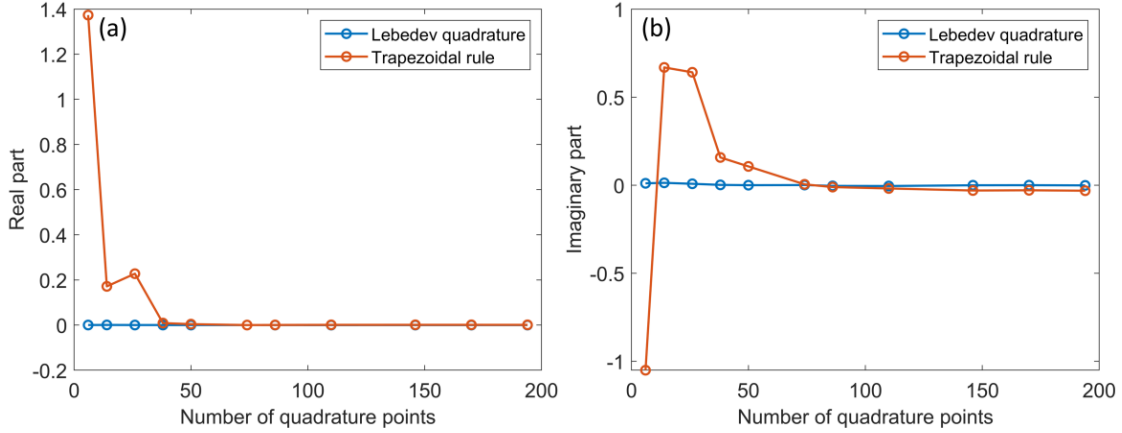

**Figure 8.** Convergence of $\frac{b_{3,-3}+b_{3,3}}{b_{3,-3}}$ of the $A_1''$ eigenmode versus the number of quadrature points. The surface integration-based numerical projection is adopted.

### 3.3. Scattering by an anisotropic nanosphere

Lastly, we use an anisotropic nanosphere to verify the accuracy of our numerical projection procedure on scatterers with optical anisotropy. The center of the nanosphere is located at the origin of coordinates and has a radius of 200 nm. The incident light is a plane wave with left-handed or right-handed circular polarization (LCP or RCP), and the incident direction is along the positive $z$ axis. The anisotropic dielectric tensor of the nanosphere is $\bar{\varepsilon}_r = \begin{pmatrix} \varepsilon_r & ig & 0 \\ -ig & \varepsilon_r & 0 \\ 0 & 0 & \varepsilon_r \end{pmatrix}$, where $\varepsilon_r = 5.2$, $g = 3$, and $i = \sqrt{-1}$. As the incident light is LCP or RCP light, the nanosphere exhibits different scattering characteristics due to its anisotropy [34]. The computation results of the scattering cross-section under LCP and RCP illumination are shown in Fig. 9. The total scattering cross-sections obtained by numerical projection are calculated by the sum of scattering cross-sections of multipoles of each order, which are in excellent agreement with the total scattering cross-sections calculated in COMSOL.

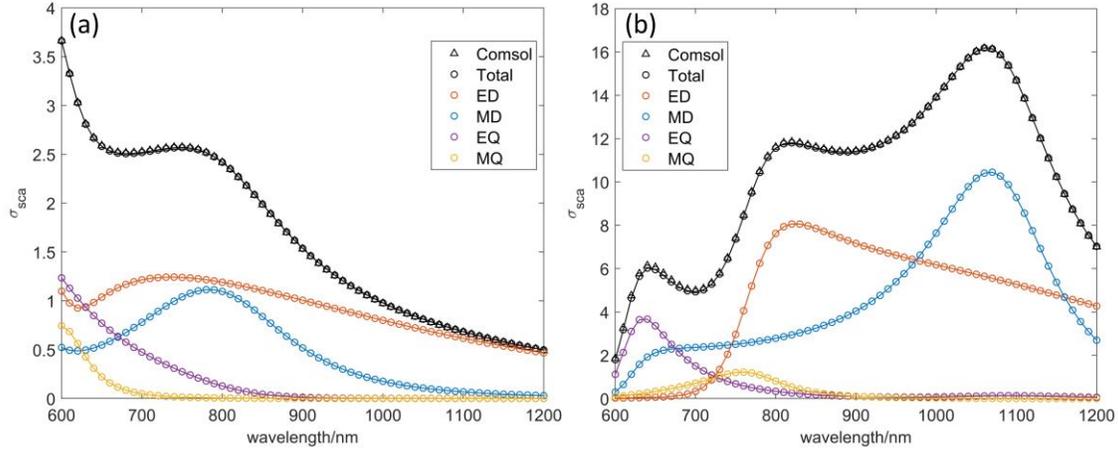

**Figure 9.** Scattering cross-sections of an anisotropic nanosphere. The incident light of subplot (a) and (b) is LCP and RCP light respectively. The triangles represent the results calculated by COMSOL, and the circles represent the results from the numerical projection algorithm.

## 4. Conclusion

In summary, we introduce the Lebedev and Gaussian quadrature methods in the numerical projection procedure of multipole decomposition for various scattering objects, and provide the corresponding open-source program, wherein the electric and magnetic multipoles up to the 8th order are coded. Two approaches for numerical projection, surface and volume integration, are reviewed. To validate our numerical projection procedure, we apply it to analyze the light scattering of an isotropic dielectric nanosphere, a symmetric scatterer, and an anisotropic nanosphere. The results show excellent agreement with those from Mie theory, symmetry constraints, and finite element simulation. Moreover, the Lebedev and Gaussian quadrature methods show faster convergence, than is achieved using the trapezoidal rule, for surface and volume integration in multipole decomposition.

Our proposed numerical projection procedure is highly efficient and accurate in analyzing the scattering properties of various nanostructures, and is a useful complement to existing works on multipole projection. Our numerical method is publicly available and could prove beneficial for those working in the field of nanophotonics, enabling convenient design of nanostructures with special scattering properties such as directional scattering, phase modulation, and high transmittance.

**Acknowledgements**

This research was funded by the National Key Research and Development Program of China (No. 2021YFB2800303), Innovation Project of Optics Valley Laboratory, and the National Natural Science Foundation of China (Grant No. 61405067).

**Author contribution** YC conceived the idea. YC and ZX and supervised the project; WG carried out the numerical simulations and analyzed the relevant data. WG and ZC wrote the manuscript. YC, ZX and WC provided revisions to the manuscript. All authors read and approved the final manuscript.

**Declarations**

The authors declare that they have no competing interests.

**Availability of data and materials**

The data that support the findings of this study are available from the corresponding author, upon reasonable request. Codes for the numerical projection procedure are available at https://github.com/HUST-CPO/Multipole-Decomposition-for-Scattering.

**References**

1. Bohren, C.F., Huffman, D.R.: *Absorption and Scattering of Light by Small Particles* (John Wiley & Sons, 1998)

2. Jackson, J.: *Classical Electrodynamics* (John Wiley & Sons, 1998), Chap. 9

3. Evlyukhin, A.B., Fischer, T., Reinhardt, C., Chichkov, B.N.: Optical theorem and multipole scattering of light by arbitrarily shaped nanoparticles. Phys. Rev. B **94**(20), 205434 (2016)

4. Terekhov, P.D., Babicheva, V.E., Baryshnikova, K.V., Shalin, A.S., Karabchevsky, A., Evlyukhin, A.B.: Multipole analysis of dielectric metasurfaces composed of nonspherical nanoparticles and lattice invisibility effect. Phys. Rev. B **99**(4), 045424 (2019)

5. Chen, W., Yang, Q., Chen, Y., Liu, W.: Scattering activities bounded by reciprocity and parity conservation. Phys. Rev. Research **2**(1), 013277 (2020)

6. Kuznetsov, A.V., Valero, A.C., Shamkhi, H.K., Terekhov, P., Ni, X., Bobrovs, V., Rybin, M.V., Shalin, A.S.: Special scattering regimes for conical all-dielectric nanoparticles. Sci. Rep. **12**(1), 21904 (2022)

7. Chen, W., Yang, Q., Chen, Y., Liu, W.: Extremize optical chiralities through polarization singularities. Phys. Rev. Lett. **126**(25), 253901 (2021)

8. Alaee, R., Rockstuhl, C., Fernandez-Corbaton, I.: Exact Multipolar Decompositions with Applications in Nanophotonics. Adv. Opt. Mater. **7**(1), 1800783 (2019)

9. Evlyukhin, A.B., Chichkov, B.N.: Multipole decompositions for directional light scattering. Phys. Rev. B **100**(12), 125415 (2019)

10. Fu, Y.H., Kuznetsov, A.I., Miroshnichenko, A.E., Yu, Y.F., Luk'yanchuk, B.: Directional visible light scattering by silicon nanoparticles. Nat. Commun. **4**(1), 1527 (2013)

11. Chen, W., Chen, Y., Liu, W.: Multipolar conversion induced subwavelength high-Q Kerker supermodes with unidirectional radiations. Laser Photonics Rev. **13**(9), 1900067 (2019)

12. van de Haar, M.A., van de Groep, J., Brenny, B., Polman, A.: Controlling magnetic and electric dipole modes in hollow dielectric nanocylinders. Opt. Express **24**(3), 2047–2064 (2016)

13. Geffrin, J.M., García-Cámara, B., Gómez-Medina, R., Albella, P., Froufe-Pérez, L.S., Eyraud, C., Litman, A., Vaillon, R., González, F., Nieto-Vesperinas, M., Sáenz, J.J., Moreno, F.: Magnetic and electric coherence in forward- and back-scattered electromagnetic waves by a single dielectric subwavelength sphere. Nat. Commun. **3**(1), 1171 (2012)

14. Chen, W., Chen, Y., Liu, W.: Singularities and Poincaré Indices of Electromagnetic Multipoles. Phys. Rev. Lett. **122**(15), 153907 (2019)

15. Kuznetsov, A.I., Miroshnichenko, A.E., Brongersma, M.L., Kivshar, Y.S., Luk'yanchuk, B.: Optically resonant dielectric nanostructures. Science **354**, aag2742 (2016)

16. Gurvitz, E.A., Ladutenko, K.S., Dergachev, P.A., Evlyukhin, A.B., Miroshnichenko, A.E., Shalin, A.S.: The High-Order Toroidal Moments and Anapole States in All-Dielectric Photonics. Laser Photonics Rev. **13**(5), 1800266 (2019)

17. Terekhov, P.D., Evlyukhin, A.B., Redka, D., Volkov, V.S., Shalin, A.S., Karabchevsky, A.: Magnetic Octupole Response of Dielectric Quadrumers. Laser Photonics Rev. **14**(4), 1900331 (2020)

18. Prokhorov, A.V., Terekhov, P.D., Gubin, M.Y., Shesterikov, A.V., Ni, X., Tuz, V.R., Evlyukhin, A.B.: Resonant light trapping via lattice-induced multipole coupling in symmetrical metasurfaces. ACS Photonics **9**(12), 3869–3875 (2022)

19. Liu, W., Zhang, J., Lei, B., Ma, H., Xie, W., Hu, H.: Ultra-directional forward scattering by individual core-shell nanoparticles. Opt. Express **22**(13), 16178-16187 (2014)

20. Liu, W.: Generalized Magnetic Mirrors. Phys. Rev. Lett. **119**(12), 123902 (2017)

21. Yang, Y., Miroshnichenko, A.E., Kostinski, S.V., Odit, M., Kapitanova, P., Qiu, M., Kivshar, Y.S.: Multimode directionality in all-dielectric metasurfaces. Phys. Rev. B **95**(16), 165426 (2017)

22. Liu, W., Miroshnichenko, A.E.: Beam Steering with Dielectric Metalattices. ACS Photonics **5**, 1733–1741 (2018)

23. Miroshnichenko, A.E., Evlyukhin, A.B., Yu, Y.F., Bakker, R.M., Chipouline, A., Kuznetsov, A.I., Luk'yanchuk, B., Chichkov, B.N., Kivshar, Y.S.: Nonradiating anapole modes in dielectric nanoparticles. Nat. Commun. **6**, 8069 (2015)

24. Saadabad, R.M., Huang, L., Evlyukhin, A.B., Miroshnichenko, A.E.: Multifaceted anapole: from physics to applications [Invited]. Opt. Mater. Express **12**, 1817-1837 (2022)

25. Khorasaninejad, M., Chen, W.T., Devlin, R.C., Oh, J., Zhu, A.Y., Capasso, F.: Metalenses at visible wavelengths: Diffraction-limited focusing and subwavelength resolution imaging. Science **352**(6290), 1190-1194 (2016)

26. Wang, S., Wu, P.C., Su, V.C., Lai, Y.C., Chen, M.K., Kuo, H.Y., Chen, B.H., Chen, Y.H., Huang, T.T., Wang, J.H., Lin, R.M., Kuan, C.H., Li, T., Wang, Z., Zhu, S., Tsai, D.P.: A broadband achromatic metalens in the visible. Nat. Nanotechnol. **13**(3), 227–232 (2018)

27. Yu, N., Capasso, F.: Flat optics with designer metasurfaces. Nat. Mater. **13**(2), 139-150 (2014)

28. Neshev, D., Aharonovich, I.: Optical metasurfaces: new generation building blocks for multi-functional optics. Light Sci. Appl. **7**(1), 58 (2018)

29. Yang, Y., Wang, W., Moitra, P., Kravchenko, I.I., Briggs, D.P., Valentine, J.: Dielectric meta-reflectarray for broadband linear polarization conversion and optical vortex generation. Nano Lett. **14**(3), 1394–1399 (2014)

30. Decker, M., Staude, I., Falkner, M., Dominguez, J., Neshev, D.N., Brener, I., Pertsch, T., Kivshar, Y.S.: High-efficiency dielectric Huygens' surfaces. Adv. Opt. Mater. **3**(6), 813-820 (2015)

31. Jahani, S., Jacob, Z.: All-dielectric metamaterials. Nat. Nanotechnol. **11**(1), 23–36 (2016)

32. Alaee, R., Rockstuhl, C., Fernandez-Corbaton, I.: An electromagnetic multipole expansion beyond the long-wavelength approximation. Opt. Commun. **407**, 17-21 (2018)

33. Evlyukhin, A.B., Reinhardt, C., Chichkov, B.N.: Multipole light scattering by nonspherical nanoparticles in the discrete dipole approximation. Phys. Rev. B **84**(23), 235429 (2011)

34. Evlyukhin, A.B., Tuz, V.R.: Electromagnetic scattering by arbitrary-shaped magnetic particles and multipole decomposition: Analytical and numerical approaches. Phys. Rev. B **107**(15) 155425 (2023)

35. Nanz, S.: *Toroidal Multipole Moments in Classical Electrodynamics: An Analysis of Their Emergence and Physical Significance* (Springer, 2016)

36. Guo, W., Cai, Z., Xiong, Z., Chen, W., Chen, Y.: HUST-CPO/Multipole-Decomposition-for-Scattering. https://github.com/HUST-CPO/Multipole-Decomposition-for-Scattering, GitHub (2023)

37. Mühlig, S., Menzel, C., Rockstuhl, C., Lederer, F.: Multipole analysis of meta-atoms. Metamaterials **5**(2), 64–73 (2011)

38. Grahn, P., Shevchenko, A., Kaivola, M.: Electromagnetic multipole theory for optical nanomaterials. New J. Phys. **14**(9), 093033 (2012)

39. Yu, J.: Symmetric gaussian quadrature formulae for tetrahedronal regions. Computer Methods in Applied Mechanics and Engineering **43**(3), 349-353 (1984)

40. Dunavant, D.A.: High degree efficient symmetrical Gaussian quadrature rules for the triangle. International Journal for Numerical Methods in Engineering **21**(6), 1129-1148 (1985)

41. Keast, P.: Moderate-degree tetrahedral quadrature formulas. Computer Methods in Applied Mechanics and Engineering **55**(3), 339-348 (1986)

42. Lebedev, V.I.: Values of the nodes and weights of ninth to seventeenth order Gauss-Markov quadrature formulae invariant under the octahedron group with inversion. USSR Computational Mathematics and Mathematical Physics **15**(1), 44-51 (1975)

43. Lebedev, V.I.: Quadratures on a sphere. USSR Computational Mathematics and Mathematical Physics **16**(2), 10-24 (1976)

44. COMSOL Multiphysics 6.0: a finite element analysis, solver and simulation software. http://www.comsol.com.

45. García-Etxarri, A., Gómez-Medina, R., Froufe-Pérez, L.S., López, C., Chantada, L., Scheffold, F., Aizpurua, J., Nieto-Vesperinas, M., Sáenz, J.J.: Strong magnetic response of submicron dielectric particles in the infrared. Opt. Express **19**(6), 4815-4826 (2011)

46. Sakoda, K.: *Optical Properties of Photonic Crystals* (Springer, 2005)

47. Xiong, Z., Yang, Q., Chen, W., Wang, Z., Xu, J., Liu, W., Chen, Y.: On the constraints of electromagnetic multipoles for symmetric scatterers: eigenmode analysis. Opt. Express **28**(3), 3073-3085 (2020)

48. Poleva, M., Frizyuk, K., Baryshnikova, K., Evlyukhin, A., Petrov, M., Bogdanov, A.: Multipolar theory of bianisotropic response of meta-atoms. Phys. Rev. B **107**(4), L041304 (2023)